# Comparison on PMT Waveform Reconstructions with JUNO Prototype

H.Q. Zhang[a,b,d], Z.M. Wang[b,d,1], Y.P. Zhang[b,d], Y.B. Huang[a,b,d], F.J. Luo[b,c], P. Zhang[a,b,d], C.C. Zhang[a,b,d], M.H. Xu[b,c], J.C. Liu[b,d], Y.K. Heng[a,b,c], C.G. Yang[a,b,d], X.S. Jiang[b,c], F. Li[b,c], M. Ye[b,c] and H.S. Chen[b,d]

[a]University of Chinese Academy of Science,
19A Yuquan Road, Shijingshan District, Beijing, China
[b]Institute of High Energy Physics, ,Chinese Academy of Sciences,
19B Yuquan Road, Shijingshan District, Beijing, China
[c]State Key Laboratory of Particle Detection and Electronics,
19B Yuquan Road, Shijingshan District, Beijing, China
[d]Key Laboratory of Particle Astrophysics,
19B Yuquan Road, Shijingshan District, Beijing, China

**Abstract**

JUNO is a 20 kton underground liquid scintillator detector aiming to determine the neutrino mass hierarchy and neutrino measurements. A prototype of JUNO is designed and set up for better understanding sub-systems of future detector. The preliminary results show that trigger rate is about 290 Hz at threshold 0.3MeV (gamma) on the surface of earth, containing cosmic muon rate ~35 Hz identified by detected energy higher than 20 MeV. For a better detector understanding with PMT signal, three waveform reconstruction algorithms are compared for PMTs with different overshoot ratios, including charge integration, waveform fitting, and deconvolution. It is concluded that the deconvolution algorithm is best to handle larger overshoot (~10%) while a selected time window length would affect the reconstruction uncertainty. Further, the three reconstruction algorithms show similar performance on consistency and uncertainty with small overshoot (~1%) waveforms and different amplitudes, which means that the simplest charge integral method is still working well for quick waveform reconstruction in this case.

## 1 Introduction

With the measurement of $\sin^2 2\theta_{13}$ by Daya Bay [1], neutrino mass hierarchy is the next novel and hot topic in particle physics. The Jiangmen Underground Neutrino Observatory (JUNO) [2] is proposed to determine the neutrino mass hierarchy using a 20 kton underground liquid scintillator (LS) detector. 20,000 large area photomultiplier tubes (PMT) [3] will be housed in JUNO to achieve the ultra-high energy resolution $3\%/\sqrt{E_{vis}}$ to measure the neutrino mass hierarchy at a confidence level of 3-4σ in 6years.

Following the schedule of JUNO, a prototype detector is designed and set up at the Institute of High Energy Physics (IHEP) Chinese Academy of Science, China. The prototype detector is designed with 51 PMTs from three companies: 8" dynode PMTs and 20" dynode PMTs from Japan

HAMAMATSU, 8" and 20" Micro Channel Plate (MCP) PMT from Nanjing Night Vision Technology Co. (NNVT) and 9" dynode PMT from Hainan Zhanchuang Photonics Co. (HZC), and more details are listed in Table 1. The motivations of JUNO prototype detector include: 1) test and study the PMT candidates in a real scintillator detector, especially for the new developed large area and high photon detection efficiency MCP PMTs; 2) test and study the new scheme of liquid scintillator and custom-developed waveform electronics; 3) study waveform analysis algorithms and detector performance.

It is reported [4] that the PMT is working with positive HV scheme where a coaxial cable supplies the HV and outputs the anode signal, and they are separated by a de-coupler. This positive HV strategy is adopted by Daya Bay[5], Borexino[6], Chooz[7], and Double Chooz [8] too. An overshoot and reflection would appear following the main pulse if the impedance is mismatched in the coupling circuit. With JUNO prototype, we have optimized the circuit to make sure the waveform with smaller overshoot for MCP-PMTs and 20'' HAMAMATSU PMTs [4]. However, the 8" HAMAMATSU tubes are potted with the HV divider by the manufacturer, which means there would be a larger overshoot around 10% as Daya bay [15]. The overshoot is a big challenge for charge measurement and system triggering which troubled Double Chooz[9], KamLAND[10], SNO[11], Borexino[12] strongly. The PMT waveforms from JUNO prototype with different overshoot ratio provide an excellent opportunity for our waveform reconstruction study, such as the reconstruction algorithm used by Daya Bay [23] and charge consistency for different amplitudes.

In this paper, the design of JUNO prototype detector, run configuration and preliminary performance will be introduced. We will show a detailed comparison on the waveform reconstruction algorithms including charge integration, waveform fitting and deconvolution algorithm.

Table 1. Detailed information of 51 tubes housed in the prototype

| Serial Number | Manufactory | Multiplier | Size/inch | Amount/piece |
| --- | --- | --- | --- | --- |
| R12860 | HAMAMATSU | Dynode | 20 | 4 |
| R5912 | HAMAMATSU | Dynode | 8 | 10 |
| XP1805 | HZC | Dynode | 9 | 11 |
| - | NNVT | MCP | 20 | 8 |
| - | NNVT | MCP | 8 | 18 |

## 2 JUNO prototype detector

The construction of JUNO prototype detector consists of LS detector, radioactivity shielding, a pure water cycling system, a custom-built electronic read-out as well as DCS system, which will be described in details in the following parts.

### 2.1 detector design and system monitoring

The prototype detector is shown in Fig.1, which re-uses the Daya Bay prototype vessel as the main container [13] with height at 208.5 cm and diameter in 198.0 cm. An acrylic sphere with 0.5m diameter locates at the center of the stainless-steel tank (SST) as a LS vessel, and is viewed by 51 PMTs dipped in pure water. The LS vessel is hold by an acrylic chimney and connected to the structure. The acrylic chimney, used for LS filling and calibration, is coaxial with the stainless steel cylindrical tank.

In the prototype, 51 tubes are housed in 5 layers, and the working tubes provide ~ 30% PMT coverage. It is worth to mention that the failed PMTs are from several sources. One is from the early potting technology used for pure water, which is confirmed recently when we un-assembly the prototype. Another reason is the early 8" MCP PMT lost its linearity response to input, and they will be replaced with the updating prototype.

The PMT overall layout is displayed on the right of Fig.1, and the location details are shown in Table 2 where the 8'' PMTs from Hamamatsu is the HQE R5912 which pre-model used in Daya Bay. A 12 layers Nano-crystalline iron film is used as the Earth magnetic field shielding (EMS) to improve the 20" PMTs' performance.

Table 2. The PMT location in the prototype

| Location-Type | 20'' HAMA /piece | 8'' HAMA /piece | 9''HZC /piece | 20'' NNVT /piece | 8'' NNVT /piece |
|---|---|---|---|---|---|
| Top | 2 | - | - | 4 | - |
| Middle-up | - | 4 | 5 | - | 7 |
| Middle-down | - | 4 | 4 | - | 8 |
| Bottom-outside | - | 2 | 2 | - | 3 |
| Bottom-inside | 2 | - | - | 4 | - |

Fig.2 is showing the radioactivity shielding system. The bottom and top are covered by 10 cm lead plus 10 cm Polypropylene (pp) plate, while the other 4 sides are shielded by 1m customized water tank made by standard stainless-steel which is widely used by water and conditioning system. One of the 4 sides is movable to allow the internal detector installation.

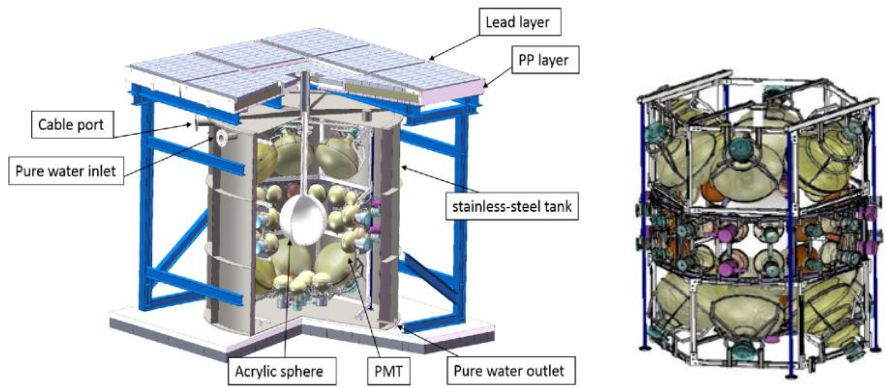

Fig.1 Left: structure of the prototype; right: the overall PMT layout consisting of four layer.

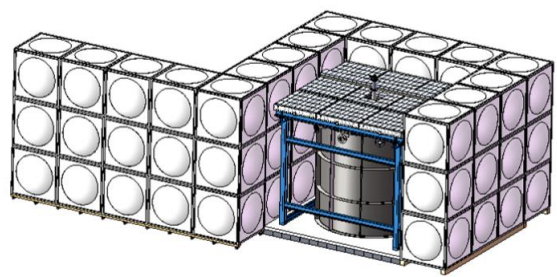

Fig.2. the layout of the customized 1m*5m*3m water tank around the prototype SST, and 10cm PP +lead 10cm layer on the top and bottom

For the JUNO prototype, there will be many materials submerged in such a small pure water tank, including stainless steel, PMT glass, cables, etc. It is a challenge to keep the water quality good in a long period without cycling system. Therefore, an ultrapure water purification and circulation system is built and the water resistivity is recorded to monitor the water quality. The produced water resistivity is around 18Mohm while the resistivity of the detector outlet is at approximately 17Mohm after 10 cycled volume expressed on the left plot of Fig.4. Additionally, nitrogen isolation is applied here to prevent ultrapure water polluted by air. The system is also used for water Radon reduction study, reported in [24].

Considering the temperature effects on dark noise of PMT and bacterial reproduction in the water, a temperature control system installed with the ultrapure water system and settle down at about 17.5 degree, lower than DYB [14]. Four sensors are used to monitor the real-time temperature in the water tank and two sensors used for the room temperature of the laboratory, and the real-time status was shown in the right one of Fig.3.

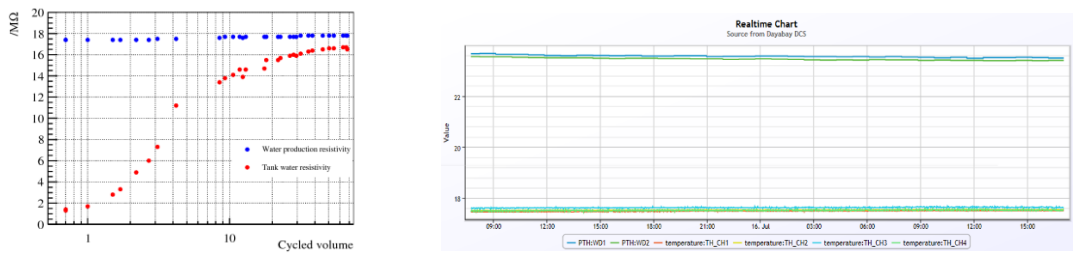

Fig.3. Left: the resistivity of water in the cycling system and water tank vs cycled volume; right: real –time temperature of room is stable at about 22.8 degree while the real –time temperature in the water tank is about 17.8 degree.

## 2.2 Electronics and Data Acquisition

PMTs in the prototype are working with positive high voltages (HV). For the electronics displayed on Fig.4, a coaxial cable supplies the HV and outputs the anode signal, which are separated by a de-coupler. The anode signal is sampled by a custom-developed Flash ADC (FADC) modules, consisting of 16 channels in each module. The 10-bit FADC operates with a sampling rate 1GHz and a maximum amplitude from 0V to 1V. The readout time window length is configurable and set to 1000 ns in our measurement to cover the baseline and scintillator photons. Three amplification scales are used with precision 0.1mV, 0.8mV and 3.2mV respectively to cover a large dynamic range. The FADC system is triggered by 8 PMTs and aims to reach ~120Hz@0.7MeV threshold.

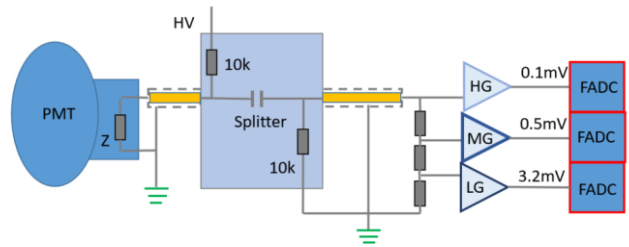

Fig.4 Schema of the integrated electronics

The data in binary from totally 112 electronics channels is taken initially and transferred to server automatically through fiber then manually converted to ROOT format for analysis. A cable map is used with the front end electronic channel ID, the electronic amplifier gear and PMT location. The data and analysis process is described in Fig.5.

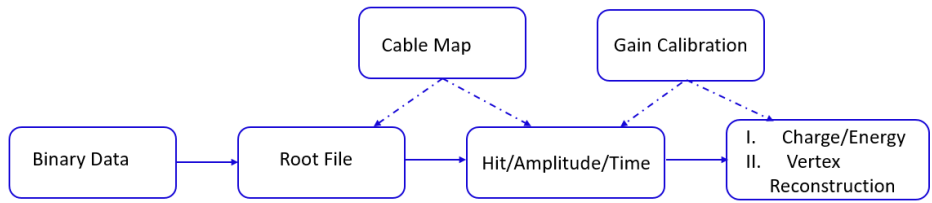

Fig.5 Data and analysis strategy

## 3. Preliminary Response

The JUNO prototype runs effectively after the installation except the mentioned MCP 8'' PMTs and potting failed tubes. During the detector calibration, the other four types of PMTs are investigated. We ran the detector several months and took the data focusing on background, gamma ray sources ($^{137}$Cs and $^{60}$Co) and neutron sources. Here, a 24 hours data on background is analyzed and shown in Fig.6, and it is concluded that the raw trigger rate is around 290 Hz with threshold ~0.3MeV gamma energy including the cosmic muon rates about 35Hz identified with threshold 20MeV.

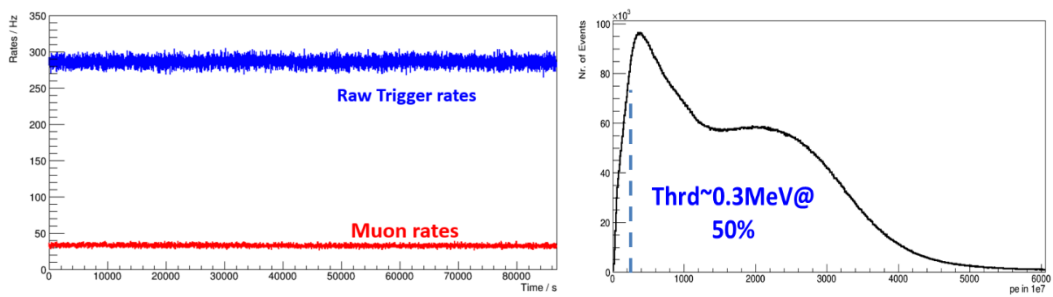

Fig.6 Left: trigger rate vs time, the red one and the blue one show the Muon rates and raw trigger rate respectively; right: the background spectrum (the dashed blue line is showing the threshold location around 0.3MeV on gamma which will be discuss more later)

A packaged $^{137}$Cs source is located in 7 positions from the bottom of LS vessel to the top of the acrylic chimney with 10cm step. Here we just of the data at the fixed location 20cm from the bottom to study the waveform reconstruction (near the center of the acrylic sphere). Based on a simple charge integration algorithm as discussed in section 4.1, the $^{137}$Cs spectrum is depicted in Fig.7 where is with the working PMT channels as mentioned and the effective threshold is around 180p.e which is got by the comparison with Geant 4 simulated spectrum.

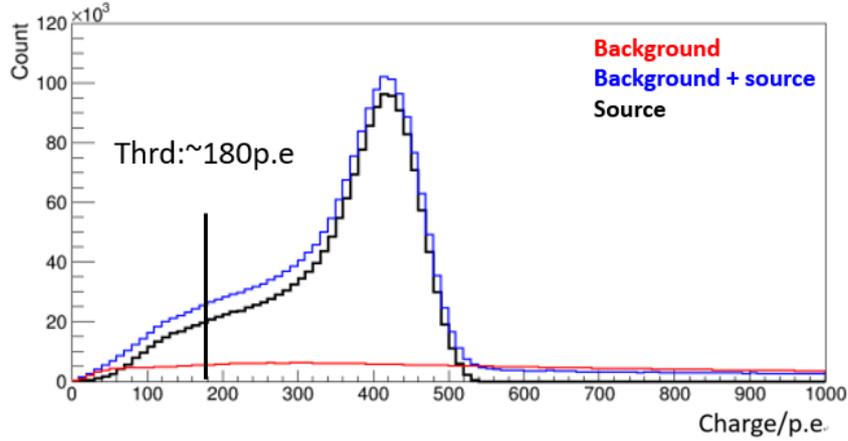

Fig.7. Measured $^{137}$Cs energy spectrum in JUNO prototype: the red curve is the background, the blue curve is the measured source spectrum and the black one is the assumed pure signal by subtracted background normalized with the counts higher than 700 p.e. The black vertical line is trying to identify the threshold location, which is ~0.3 MeV according to $^{137}$Cs response.

## 4 Waveform Reconstruction

As described, the custom-developed FADC samples the raw waveforms for all the PMTs, while no bias waveform reconstruction is a big challenge to estimate the signal charge especially for waveforms with large overshoot, which will further contribute to the energy response linearity. The overshoot of PMT waveform in Daya Bay is studied by deconvolution algorithm [23] and the overshoot of PMT waveform in JUNO has been controlled to ~1%. It is valuable to study an effective algorithm to guarantee the waveform reconstruction quality considering the costs and to simplify the analysis with a real detector. The waveforms with large (10%) and small (1%) overshoot in JUNO prototype provide a perfect opportunity to validate the charge reconstruction algorithms. In this paper, we studied the three different PMT waveform reconstruction algorithms including the charge integration (CI), waveform fitting (WF) and deconvolution algorithm (DA), aiming to develop and validate the charge reconstruction approach. The LED and $^{137}$Cs radioactive source calibration data are analyzed here for the discussion where the saturated waveforms are removed already.

**4.1 Charge Integration (CI)**
The simplest and common method to calculate the PMT anode output charge is to integrate the waveform samples in a window, where the integration window is possible to contribute to the final charge calculation uncertainty. To minimize the contribution from integration window relative to a single signal location, a procedure to determine the integration window location is developed according to the peak location of each waveform, which comprised three steps and displayed in Fig.8:

Step 1, Maximum peak location determination: Average Waveform Superimposed. Because of the waveform readout window is related to the common trigger of our electronics, which can't be stable enough relative each PMT waveform because of time of flying for different location and sources, PMT TTS. Each PMT/channel should superimpose to its own waveforms to obtain its maximum peak location through the averaged waveform.

Step 2, Peak finding Window determination. Take the maximum peak location from step 1 as reference for each PMT/channel, and define a 60 ns front and 60ns after this reference as peak finding window to find the waveform peak location for each waveform.

Step 3, Final Integration Window and charge calcualation. The peak location for each waveform is found in the peak finding window from step 2. Then the waveform is integrated in 20ns front and 55ns (or 80ns and 100ns as discussed later) after the peak location to get the integrated charge, where the baseline of each waveform also is calculated according to the peak location (-100ns, -50ns to peak location) and subtracted from the charge integration.

Following the listed steps, charge of each waveform can be calculated. Here showing an example with a 8'' HAMAMATUS tubes' waveform with large overshoot (right of Fig.8), the results with three different integration windows, respectively 75 ns (front 20 ns + after 55 ns), 100 ns (20+80 ns) and 120 ns (20+100 ns) shown in the left of Fig.8, will be discussed further in section 4.4.

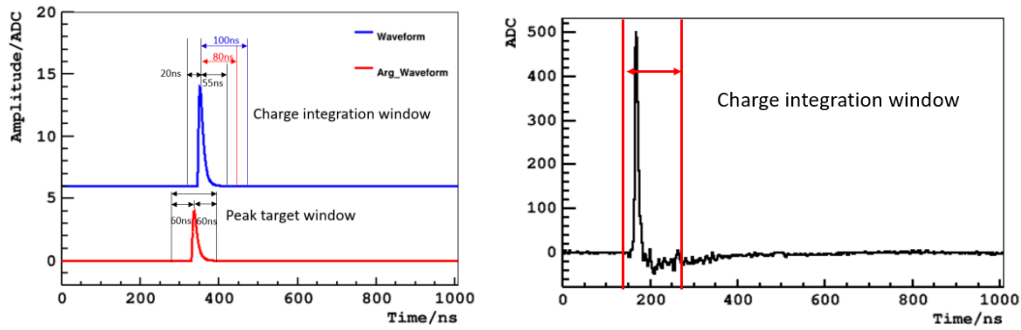

Fig.8 Left: the definition of charge integration window (the average waveform and single waveform are generated by toy MC as example); right: measured waveform and integration window example.

**4.2 Waveform Fitting (WF)**

To better understand the overshoot, a waveform fitting is studied. A waveform model with asymmetric shape, displayed in Equation (3.1)[15], is suitable for the PMT output waveform if there is not a large overshoot.

However, the de-coupler used to split the high-voltage and the pulse signal would result in baseline distortion after the main pulse, which is known as overshoot. Concerning the overshoot, we follow [16] to have an exponential tail (Equation 3.2) and a Gaussian (Equation 3.3), which correspond to the tail and main-part of the overshoot. And Equation 3.4 is proposed to fit all the waveforms including with large overshoot. The parameters and validated range of the fitting are listed in Table 3, where the initial values are extracted from each waveform and the overshoot parameters are referring to Daya Bay [15].

The fitting window is configurable, and we compared three different windows which are same with the integration windows as listed in Section 4.1. Actually, the fitting is just a waveform smoothing rather than a physics fitting to photoelectrons and hit time, that means it still needs to use another integration to calculate the charge. Here we will use "Fit+Integral" to label the waveform fitting for comparison. A fitting example is depicted in Fig.9 and results will be discussed in section 4.4.

$$U_{peak}(t) = U_B + U_0 \cdot \exp\left(-\frac{1}{2}\left(\frac{In(\frac{t-t_0}{\tau_0})}{\sigma_0}\right)^2\right) \quad (3.1)$$

$$U_{os1}(t) = U_1 \cdot \frac{1}{\exp(\frac{t_1-t}{10ns})+1} \cdot \exp\left(-\frac{t}{\tau_{os}}\right) \quad (3.2)$$

$$U_{os2}(t) = U_2 \cdot \exp\left(-\frac{1}{2}\left(\frac{t-t_2}{\sigma_2}\right)^2\right) \quad (3.3)$$

$$U(t) = U_{peak} - U_{os1} - U_{os2} \quad (3.4)$$

Table 3. Parameters definition and initial value for waveform fitting ("-"means no fixed value and no adjustable range in the fitting, and they are determined by the waveform pretreatment: automatic peak finding, baseline estimation and etc.)

| Serial No | Parameter | Setting Value | Adjustable Range | Meaning | Description | Equation |
|---|---|---|---|---|---|---|
| 0 | $U_0$ | - | (5,1000)ADC | Amplitude | Main-pulse | eq. 3.1 |
| 1 | $\tau_0$ | - | (4,20) | width | Main-pulse | eq. 3.1 |
| 2 | $\sigma_0$ | - | (0.2,0.8) | Shape factor | Main-pulse | eq. 3.1 |
| 3 | $t_0$ | - | - | Peak position | Main-pulse | eq. 3.1 |
| 4 | $U_B$ | - | - | Baseline offset | Main-pulse | eq. 3.1 |
| 5 | $t_1$ | - | - | Starting time | Overshoot | eq.3.2 |
| 6 | $\tau_{os}$ | 150 | (0,300) | Decay time | Overshoot | eq.3.2 |
| 7 | $t_2$ | - | - | Peak position | Overshoot | eq. 3.3 |
| 8 | $\sigma_2$ | 20 | (10,50) | width | Overshoot | eq. 3.3 |

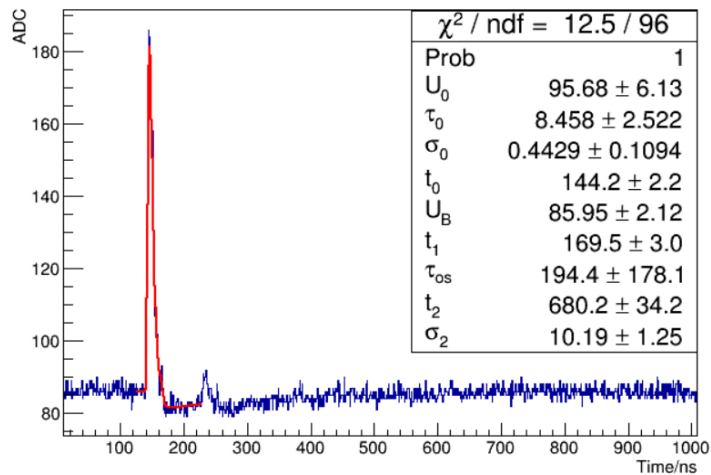

Fig.9 A fitting example of overshoot waveform

### 4.3 Deconvolution Algorithm (DA)

Deconvolution is a robust tool in Digital signal processing (DSP) [17]. It is widely used in the fields of channel equalization [18], image restoration [19], speech recognition [20], seismology [21], non-destructive testing [22] etc. In particle physics, deconvolution also is a powerful tool for PMT waveform analysis such as charge reconstruction, where frequency domain will avoid some issues in time domain especially with high precision sampled waveform [23]. Theoretically, the PMT signal is convolved from the electronic noise and PMT response. Here we will reverse the PMT signal process following the response: PMT single photoelectron response (SPE) model building, filtering, Fast Fourier Transform (FFT, realized by CERN ROOT package), de-convolution, inverse FFT (realized by CERN ROOT package) and pulse integration in time domain to get charge.

A SPE template, calculated from 10,000 SPE waveforms of a single PMT, will be used in the waveform deconvolution algorithm, where the SPE waveform charge is limited in (0.9, 1.1) p.e. which is calculated with integration window method and the pulses' FWHM is required larger than 3 ns to filter noise and make the SPE template more clean in frequency domain. Fig.10 is showing the template examples: left is the averaged SPE waveform in time domain and right is its FFT converted spectrum in frequency domain, which will be used for charge calculation.

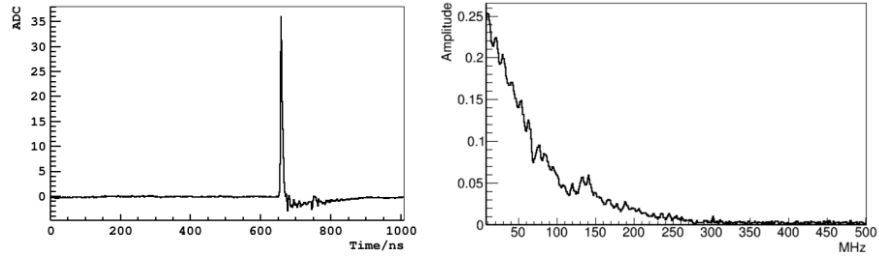

Fig.10 A SPE template in time domain (left) and in frequency domain (right) averaged from 10000 waveforms.

In frequency domain, a custom-build low-band pass filter will be applied as Equation 3.5 to remove noise from time domain, where the selected parameters $\mu, b, \sigma$ would affect the model performance. Here ($\mu = 125, b = 375, \sigma = 20$) is selected based on tests.

For a measured PMT waveform in time domain (left on Fig. 11), without amplitude saturation, it will be converted into frequency domain firstly by FFT, then multiplied by the Equation 3.5 filter and divided by SPE frequency template on each bin. Next, the inverse FFT is employed to convert the spectrum in frequency domain back to the time domain, named pulse spectrum (right on Fig. 11), which has already normalized to SPE template and can be considered as charge spectrum directly in p.e.. Following the same procedure as in section 4.1 for charge integration, the charge in p.e. can be calculated relative to different target window. It is normal that ringing shows near the peaks in the pulse spectrum as right of Fig. 11, which is considered as the Gibbs effects [23] from the high frequency filter. To make the procedure more clear, a schematic diagram of the deconvolution algorithm are displayed in Fig.12, and more results will be discussed in Section 4.4.

$$filter = \begin{cases} 1, x < \mu \\ e^{-\frac{(x-u)^2}{2\sigma^2}}, \mu \leq x \leq b \\ 0, x > b \end{cases} \quad (3.5)$$

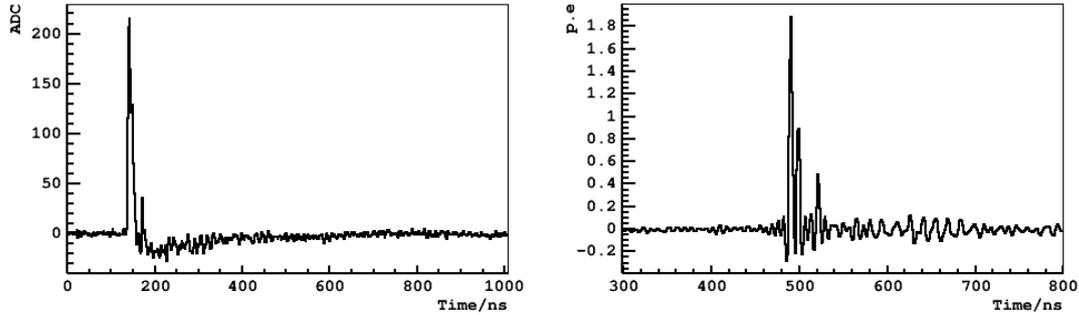

Fig.11. Left: a raw waveform with overshoot; right: the re-converted pulse spectrum after deconvolution algorithm.

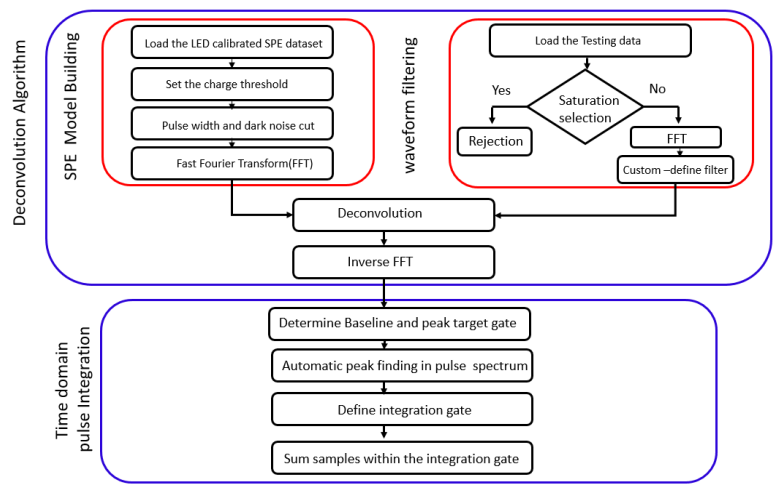

Fig.12 Schematic diagram of deconvolution algorithm: SPE Model Building, waveform filtering, Deconvolution, Inverse FFT and Pulse integration.

**4.4 Reconstruction Results**

Waveform reconstructions with two different overshoot ratios are studied using the three waveform reconstruction algorithms, where waveforms with over 10% overshoot ratio are from 8'' HAMAMATSU PMTs and waveforms with 1% overshoot are from 20'' HAMAMATSU PMTs in JUNO prototype.

**4.4.1 Large overshoot (~10%)**

The data are used for the comparison: a calibration spectrum with $^{137}$Cs at the fixed location 20cm from the bottom and one LED calibration sets with different light intensities covering from single photoelectron to hundreds of photoelectrons. It is compared among the three algorithms with the same data. For a single reconstruction algorithm, three integration window lengths are checked on the stability and consistency of the estimated energy.

A parameter, "uncertainty", is defined here to compare the consistency among the three integration windows: the reconstructed charge RMS from the 3 windows where the charge from the minimum window (-20 ns, +55 ns) is defined as the reference. An example with $^{137}$Cs data is shown. The left of Fig. 13 shows the reconstructed charge and Table 4 shows the uncertainty of the three reconstruction algorithms among different integration windows: waveform fitting (10.92%)> charge integration (9.00%)>Deconvolution Algorithm (0.72%), which means that Deconvolution Algorithm is more stable for different integration window.

The integration window dependence for the three algorithms is checked further with different LED intensities' data. The defined uncertainties of the three algorithms are plotted on right of Fig.13, where the deconvolution algorithm still has minimum uncertainty for all points, while the waveform fitting and charge integration shows larger bias for the waveforms with large overshoot.

Table 4 Reconstruction results of waveforms with 10% overshoot vs. the integration window (data with $^{137}$Cs), and the results from window (-20 ns, +55 ns) is selected as the reference.

| Integration Gate/ Fitting Gate (ns) | (-20,+55) Mean/Error | (-20,+80) Mean/Error | (-20,+100) Mean/Error | RMS | RMS/(-20,+55) /%(uncertainty) |
|---|---|---|---|---|---|
| CI (p.e) | 34.67/0.18 | 30.54/0.15 | 27.04/0.14 | 3.12 | 9.00 |
| WF(p.e) | 33.02/0.17 | 27.66/0.15 | 24.26/0.12 | 3.61 | 10.92 |
| DA(p.e) | 33.18/0.17 | 33.65/0.17 | 33.72/0.17 | 0.24 | 0.72 |

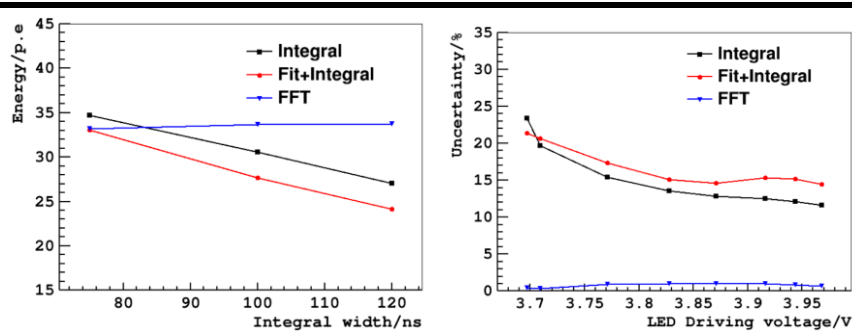

Fig.13 Left: $^{137}$Cs calibration data, the reconstructed energies with error bar (too small to be seen here) with integration window width 75ns, 100ns and 120ns and the three charge reconstruction algorithms with 10% overshoot; Right: uncertainties of the three algorithms vs. LED driving voltage (light intensities).

**4.4.2 Small overshoot (1%)**

The same data of $^{137}$Cs and LED calibration are used to investigate the waveform with small (1%) overshoot. The analysis and definition are the same as used in section 4.4.1. The uncertainties of charge integration, waveform fitting and deconvolution algorithm all are less than 2% for $^{137}$Cs data as listed in Table 5 and the reconstructed average energy with error bar in different integration windows are displayed on left of Fig.14. The integration window has little effect too on the uncertainties along different intensities checked with LED, and the plot on the right of Fig. 14 reveals that these three algorithms all meet the requirements in 2% uncertainty, even the deconvolution algorithm has better results.

Table 5 Reconstruction results of waveforms with 1% overshoot vs. the integration window (data of $^{137}$Cs), and the reconstruction result with window (-20 ns, +55ns) is selected as reference

| Integration Gate/ Fitting Gate (ns) | (-20,+55) Mean/Error | (-20,+80) Mean/Error | (-20,+100) Mean/Error | RMS | RMS/(-20,+55) /%(uncertainty) |
|---|---|---|---|---|---|
| CI (p.e) | 108.20/0.47 | 110.50/0.47 | 113.40/0.48 | 2.13 | 1.97 |
| WF(p.e) | 104.50/0.47 | 105.90/0.47 | 107.80/0.48 | 1.35 | 1.29 |
| DA(p.e) | 106.60/0.49 | 107.10/0.46 | 107.80/0.46 | 0.49 | 0.46 |

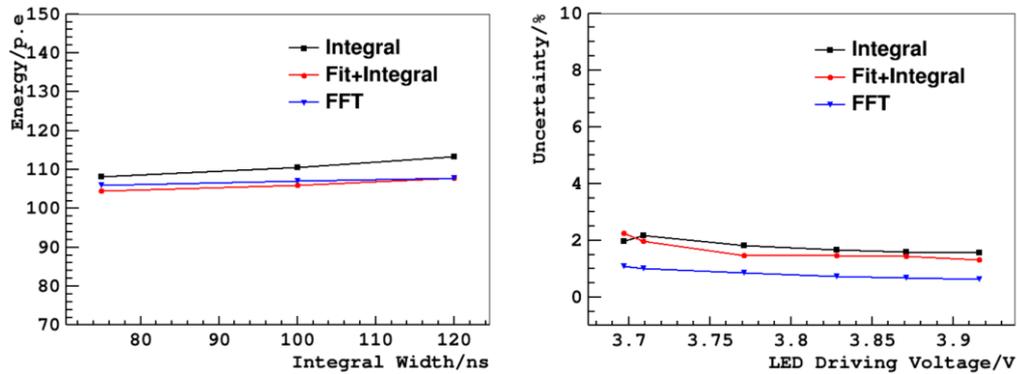

Fig.14 Left: $^{137}$Cs calibration data, the reconstructed energies with error bar (too small to be seen here) with integration window width 75ns, 100ns and 120ns and the three charge reconstruction algorithms with 1% overshoot; Right: uncertainties of the three algorithms vs. LED driving voltage (light intensities).

Till now, we have compared the integration window effects on the reconstruction uncertainty and consistency for the three algorithms: charge integration, waveform fitting and deconvolution algorithm, furthermore, we will discuss the difference among the algorithms with the same integration window.

Here, "Relative Difference" is defined to compare the difference among these three algorithms with the same given integration window: difference to the average of the 3 results. And "Self Difference" is defined as the distribution of "Relative Difference" of different light intensities with the same integration window. The results of "Relative Difference" and "Self Difference" are shown in Table 6 and left of Fig.15 with integration window (-20 ns, +55 ns). There is no obvious difference of the three algorithms for small overshoot (1%) waveform.

For integration window length of 100ns and 120ns, the results of "Self Difference" are also in 1% level as shown on right of Fig.15.

Table 6. The "relative difference" of charge Integration, waveform fitting and deconvolution algorithm vs. the LED driving voltage from 3.697V to 3.916V and the "self difference" ( with 1% overshoot waveform, with integration window length 75ns)

| LED Drving HV/V | FFT Reconstrction Charge/p.e | Fitting Reconstrction Charge/p.e | Integration Reconstrction Charge/p.e | Average Value/p.e | FFT Relative Difference /% | Fitting Relative Difference /% | Integration Relative Difference /% |
|---|---|---|---|---|---|---|---|
| 3.697 | 4.53 | 4.60 | 4.66 | 4.60 | -1.51% | 0.10% | 1.41% |
| 3.709 | 7.70 | 7.87 | 7.99 | 7.85 | -1.94% | 0.15% | 1.79% |
| 3.771 | 22.65 | 23.03 | 23.39 | 23.02 | -1.62% | 0.03% | 1.59% |
| 3.828 | 58.41 | 59.71 | 60.21 | 59.44 | -1.74% | 0.45% | 1.29% |
| 3.871 | 115.90 | 118.60 | 119.4 | 117.97 | -1.75% | 0.54% | 1.22% |
| 3.916 | 222.60 | 227.40 | 229.3 | 226.43 | -1.69% | 0.43% | 1.27% |
| | | **Self Difference** | | | 0.43% | 0.44% | 0.57% |

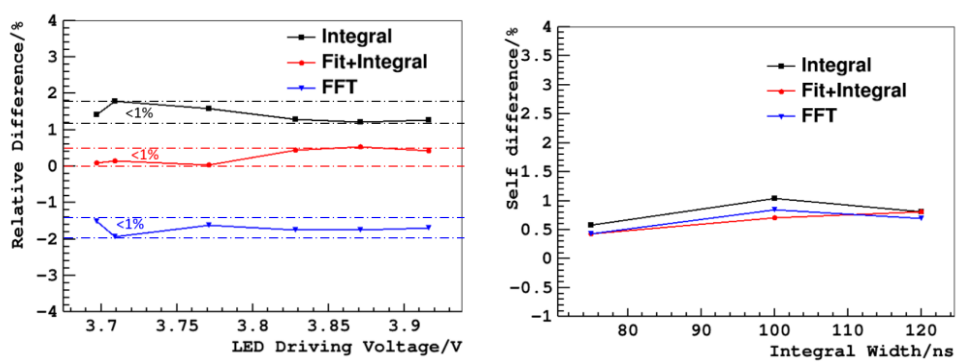

Fig.15 Left: relative difference vs. LED driving voltage in 1% overshoot waveform with integration window length 75ns. (The interval between black dotted lines indicates the "Self Difference" of charge integration) Right: "Self Difference" vs the integration window length.

## 5. Conclusion and Discussion

In this paper, we set up a multi-purpose JUNO prototype detector consisting of most sub-systems. The prototype has been run since 2016 and its preliminary response meets our expectation. The results show that raw trigger rate is about 290 Hz at gamma threshold 0.3MeV on the surface of earth, containing cosmic muon rate ~35 Hz at 20MeV threshold. In the waveform reconstruction algorithms' comparison, deconvolution algorithm would be the most effective method for the waveforms with large overshoot (10%) ratio since the integration window length has little influence on it. Following the results for 1% overshoot ratio waveforms, we suggest that the simplest waveform integration can be considered for future fast preliminary reconstruction because the three algorithms give the same level results and considering that the setting of parameters, running and failure rate of waveform fitting is a big challenge, and it needs more work to build SPE model for huge number PMTs for deconvolution algorithm.

## Acknowledgements

This work is supported by the National Natural Science Foundation of China (Grant No. 11875282), and the Strategic Priority Research Program of the Chinese Academy of Sciences (Grant No. XDA10010200, XDA10010300). We are extremely thanks to all who helps to make it is possible on the detector installation and running.